\documentclass[prd,aps,nofootinbib,superscriptaddress,floatfix,preprintnumbers,twocolumn]{revtex4}
\usepackage{hyperref,amssymb,amsmath,graphicx,xcolor,slashed}

\DeclareMathOperator\Tr{Tr}

\begin{document}
\preprint{MSUHEP-20-019}
\pacs{12.38.-t, 
      11.15.Ha, 
      12.38.Gc  
}

\title{Lattice Nucleon Isovector Unpolarized Parton Distribution in the Physical-Continuum Limit}

\author{Huey-Wen Lin}
\email{hwlin@pa.msu.edu}
\affiliation{Department of Physics and Astronomy, Michigan State University, East Lansing, MI 48824}
\affiliation{Department of Computational Mathematics,
  Science and Engineering, Michigan State University, East Lansing, MI 48824}

\author{Jiunn-Wei Chen}
\email{jwc@phys.ntu.edu.tw}
\affiliation{Department of Physics, Center for Theoretical Physics, and Leung Center for Cosmology and Particle Astrophysics, National Taiwan University, Taipei, Taiwan 106}

\author{Rui Zhang}
\email{zhangr60@msu.edu}
\affiliation{Department of Physics and Astronomy, Michigan State University, East Lansing, MI 48824}
\affiliation{Department of Computational Mathematics,
  Science and Engineering, Michigan State University, East Lansing, MI 48824}

\begin{abstract}
We present the first lattice-QCD calculation of the nucleon isovector unpolarized parton distribution functions (PDFs) at the physical-continuum limit using Large-Momentum Effective Theory (LaMET).
The lattice results are calculated using ensembles with multiple sea pion masses with the lightest one around 135~MeV, 3 lattice spacings $a\in[0.06,0.12]$~fm, and multiple volumes with $M_\pi L$ ranging 3.3 to 5.5.
We perform a simultaneous chiral-continuum extrapolation to obtain RI/MOM renormalized nucleon matrix elements with various Wilson-link displacements in the continuum limit at physical pion mass.
Then, we apply one-loop perturbative matching to the quasi-PDFs to obtain the lightcone PDFs.
We find the lattice-spacing dependence to be much larger than the dependence on pion mass and lattice volume for these LaMET matrix elements.  Our physical-continuum limit unpolarized isovector nucleon PDFs are found to be consistent with global-PDF results.
\end{abstract}

\maketitle

\section{Introduction}

Precision determination of parton distribution functions (PDFs) is
not only important to probing unknowns of the Standard Model but also to advance interpretation of high-energy experiments searching for signs of physics beyond the Standard Model.
In addition to energy-frontier experiments like the LHC, there are also many mid-energy experimental efforts around the world, such as at Brookhaven and Jefferson Laboratory in the United States,  GSI in Germany, J-PARC in Japan, or a future electron-ion collider (EIC).
These are set to explore the less-known kinematic regions of nucleon structure and more.
The pursuit of PDFs has led to collaborations of theorists and experimentalists working side-by-side to take advantage of all available data, evaluating different combinations of input theories, parameter choices and assumptions, resulting in multiple global-PDF determinations.
Comparison of these different global-fit determinations of the PDFs is important to reveal hidden uncertainties in PDF data sets.
Often, in kinematic regions where experimental data are plentiful or overconstrained, such as the mid-$x$ region of the PDFs, there is consistency among different PDF sets.
However, in the regions where data are sparse or suffer from complicated nuclear effects, such as at small- or large-$x$ or for heavy-flavor PDFs, disagreements are seen.
For more details, we refer readers to the non-technical review in Ref.~\cite{Lin:2017snn}.
A nonperturbative approach from first principles, such as lattice QCD (LQCD), can provide the necessary inputs to fill gaps in the experimental data or add information to constrain global fits.
For a long while, lattice PDF calculations were limited to moments only, that is, where the $x$ dependence of the PDF is integrated out.
Precision lattice determinations of the moments (after removing all lattice artifacts, such as discretization and finite-volume effects) can have significant impact on determinations of the PDFs~\cite{Lin:2017stx,Lin:2017snn}.

Large-momentum effective theory (LaMET)~\cite{Ji:2013dva,Ji:2014gla} enables computation of the Bjorken-$x$ dependence of hadron PDFs on a Euclidean lattice.
LaMET relates equal-time spatial correlators, whose Fourier transforms are called quasi-PDFs, to PDFs in the limit of infinite hadron momentum.
For large but finite momenta accessible on a realistic lattice, LaMET relates quasi-PDFs to physical ones through a factorization theorem, the proof of which was developed in Refs.~\cite{Ma:2017pxb,Izubuchi:2018srq,Liu:2019urm}.
Since LaMET was proposed, a lot of progress has been made in the theoretical understanding of the formalism~\cite{Xiong:2013bka,Ji:2015jwa,Ji:2015qla,Xiong:2015nua,Ji:2017rah,Monahan:2017hpu,Stewart:2017tvs,Constantinou:2017sej,Green:2017xeu,Izubuchi:2018srq,Xiong:2017jtn,Wang:2017qyg,Wang:2017eel,Xu:2018mpf,Chen:2016utp,Zhang:2017bzy,Ishikawa:2016znu,Chen:2016fxx,Ji:2017oey,Ishikawa:2017faj,Chen:2017mzz,Alexandrou:2017huk,Constantinou:2017sej,Green:2017xeu,Chen:2017mzz,Chen:2017mie,Lin:2017ani,Chen:2017lnm,Li:2016amo,Monahan:2016bvm,Radyushkin:2016hsy,Rossi:2017muf,Carlson:2017gpk,Ji:2017rah,Briceno:2018lfj,Hobbs:2017xtq,Jia:2017uul,Xu:2018eii,Jia:2018qee,Spanoudes:2018zya,Rossi:2018zkn,Liu:2018uuj,Ji:2018waw,Bhattacharya:2018zxi,Radyushkin:2018nbf,Zhang:2018diq,Li:2018tpe,Braun:2018brg,Detmold:2019ghl,Sufian:2020vzb,Shugert:2020tgq,Green:2020xco,Braun:2020ymy,Lin:2020ijm,Bhat:2020ktg,Chen:2020arf,Ji:2020baz,Chen:2020iqi,Chen:2020ody,Alexandrou:2020tqq,Fan:2020nzz,Ji:2020brr}.
The method has been applied in lattice calculations of PDFs for the up and down quark content of the  nucleon~\cite{Lin:2014zya,Chen:2016utp,Lin:2017ani,Alexandrou:2015rja,Alexandrou:2016jqi,Alexandrou:2017huk,Chen:2017mzz,Lin:2018pvv,Alexandrou:2018pbm,Chen:2018xof,Alexandrou:2018eet,Lin:2018qky,Liu:2018hxv,Wang:2019tgg,Lin:2019ocg,Liu:2020okp,Lin:2019ocg,Zhang:2019qiq,Alexandrou:2020qtt},
$\pi$~\cite{Chen:2018fwa,Izubuchi:2019lyk,Lin:2020ssv,Gao:2020ito} and $K$~\cite{Lin:2020ssv} mesons,
and the $\Delta^+$~\cite{Chai:2020nxw} baryon.
Despite limited volumes and relatively coarse lattice spacings, previous state-of-the-art nucleon isovector quark PDFs, determined from lattice data at the physical pion mass, have shown reasonable agreement~\cite{Lin:2018pvv,Alexandrou:2018pbm} with phenomenological results extracted from the experimental data.
Encouraged by this success, LaMET has also been extended to twist-three PDFs~\cite{Bhattacharya:2020cen,Bhattacharya:2020xlt,Bhattacharya:2020jfj},
as well as gluon \cite{Fan:2018dxu,Fan:2020cpa},
strange and charm distributions~\cite{Zhang:2020dkn}.
It was also applied to meson distribution amplitudes~\cite{Zhang:2017bzy,Chen:2017gck,Zhang:2020gaj,Hua:2020gnw}
and generalized parton distributions (GPDs)~\cite{Chen:2019lcm,Alexandrou:2020zbe,Lin:2020rxa,Alexandrou:2019lfo}.
Attempts have also been made to generalize LaMET to transverse momentum dependent (TMD) PDFs~\cite{Ji:2014hxa,Ji:2018hvs,Ebert:2018gzl,Ebert:2019okf,Ebert:2019tvc,Ji:2019sxk,Ji:2019ewn,Ebert:2020gxr},
to calculate the nonperturbative Collins-Soper evolution kernel~\cite{Ebert:2018gzl,Shanahan:2019zcq,Shanahan:2020zxr}
and soft functions~\cite{Zhang:2020dbb} on the lattice.
LaMET also brought renewed interest in earlier approaches~\cite{Liu:1993cv,Detmold:2005gg,Braun:2007wv,Bali:2017gfr,Bali:2018spj,Detmold:2018kwu,Liang:2019frk}
and inspired new ones~\cite{Ma:2014jla,Ma:2014jga,Chambers:2017dov,Radyushkin:2017cyf,Orginos:2017kos,Radyushkin:2017lvu,Radyushkin:2018cvn,Zhang:2018ggy,Karpie:2018zaz,Joo:2019jct,Radyushkin:2019owq,Joo:2019bzr,Balitsky:2019krf,Radyushkin:2019mye,Joo:2020spy,Can:2020sxc}.
For recent reviews on these topics, we refer readers to Refs.~\cite{Lin:2017snn,Cichy:2018mum,Zhao:2020vll,Ji:2020ect,Ji:2020byp} for more details.

To further improve the lattice computations at physical pion mass, the remaining lattice systematics must be treated, by extrapolation to infinite volume and the continuum limit.
This is a critical next step to create a lattice PDF calculation with fully controlled systematics. 
Since our calculation uses the quasi-PDF method, we consider only quasi-PDF results for comparison of systematic uncertainty.
The first study of finite-volume systematics was done in Ref.~\cite{Lin:2019ocg} with isovector both polarized and unpolarized nucleon PDFs;
three lattice volumes (2.88, 3.84, 4.8~fm) were studied at pion mass 220-MeV and nucleon momenta 1.3 and 2.6~GeV, and no noticeable finite-volume dependence was found.
This is consistent with a later study in chiral perturbation theory (ChPT)~\cite{Liu:2020krc}, which showed that momentum boost reduces the finite-volume effect, since the length contraction of the hadron makes the lattice effectively bigger.
ChPT also showed that for nucleon momenta greater than 1~GeV and the lattice size times pion mass greater than 3, then the finite-volume effect on the isovector nucleon PDF is less than $1\%$.
This conclusion is consistent with the numerical findings of Ref.~\cite{Lin:2019ocg} and suggests that the finite-volume effect is negligible at current lattice precision.

Continuum extrapolation is also important for LaMET due to potential operator mixing in the nonlocal operators. 
The nonlocal operators for the quasi-PDFs can mix with a tower of higher-dimensional operators at $\mathcal{O}(a)$, even if all symmetries (including chiral symmetry) are restored~\cite{Green:2017xeu,Chen:2017mie,Green:2020xco}.
This is different from the situation for local operators, where mixing can occur at $\mathcal{O}(a^2)$ if a chiral lattice fermion action is used.
To ensure that such operator mixings do not contaminate the final results of the lattice PDF calculations, it is important to take the continuum limit.
There have been some studies of the continuum extrapolation of the quasi-PDF method in the pion and kaon distribution amplitudes~\cite{Zhang:2020gaj} and in nucleon PDFs~\cite{Alexandrou:2020qtt};
both cases use three lattice spacings but a single heavy quark mass with $M_\pi > 300$~MeV.
Ref.~\cite{Lin:2020ssv} determines valence-quark PDFs of the pion and kaon using two lattice spacings (0.06 and 0.12~fm) and 3 pion masses ($M_\pi \in [220,690]$~MeV).
This work is the first study of lattice PDFs to take the continuum-physical limit of the matrix elements with a sufficient number of lattice spacings and light pion masses, an important step toward precision PDFs from lattice QCD.

\section{Lattice Parameters and Setup}\label{sec:set_up}

In this work, we use clover lattice fermion action for the valence quarks on top of 2+1+1 flavors (degenerate up and down quarks plus strange and charm quarks at their physical masses in the QCD vacuum) of hypercubic (HYP)-smeared~\cite{Hasenfratz:2001hp} highly improved staggered quarks (HISQ)~\cite{Follana:2006rc,Bazavov:2012xda} in configurations generated by MILC Collaboration.
The lattice parameters include lattice spacings $a \in [0.06,0.12]$~fm, pion mass $M_\pi \in [135,318]$~MeV and box size $L\in [2.9,5.5]$~fm (which make $M_\pi L\in [3.3,5.5]$).
The quark masses for the clover fermions have been tuned to reproduce the lightest sea staggered pseudoscalar meson masses for the light and strange quarks, and the clover parameters are set to the tree-level tadpole-improved values.
This setup is the same as the one used in works done by PNDME Collaboration in many studies of nucleon structure~\cite{Gupta:2018qil,Bhattacharya:2015wna,Bhattacharya:2015esa,Bhattacharya:2013ehc}. 
Note that any mixed-action approach results in a non-unitary lattice-QCD formulation with the possibility of exceptional configurations.
Signatures of such configurations, which manifest at sufficiently small quark mass, include correlation functions with anomalously large values that bias the ensemble average, and failure of the clover Dirac matrix solver to converge due to poor condition number. 
These two signatures have been observed at $a\approx 0.15$~fm and 0.12~fm at $M_\pi \approx 135$~MeV, and these ensembles are excluded from use in mixed-action calculations.
The other ensembles are carefully checked for the relevant signatures, and exceptional configurations are absent for the $M_\pi\in \{220, 310\}$~MeV MILC ensembles~\cite{Bazavov:2012xda} at 0.12~fm and finer lattice spacings, as well as for 0.09 and 0.06~fm near the physical pion mass.
There are no issues that we have observed for any observable on the ensembles used in this calculation.

For the nucleon matrix-element measurement, we use Gaussian momentum smearing~\cite{Bali:2016lva} for the quark field
\begin{multline}
\psi(x) \rightarrow S_\text{mom}\psi(x) = \\
\frac{1}{1+6\alpha} \Big(\psi(x)
+ \alpha \sum_j U_j(x)e^{ik\hat{e}_j}\psi(x+\hat{e}_j)\Big)\,,
\label{eq:moms}
\end{multline}
where $k$ is the momentum-smearing parameter, which can be tuned separately on each ensemble for optimal signal-to-noise ratios in the matrix elements of the desired nucleon boost momentum.
$U_j(x)$ are the gauge links in the $j$ direction, and $\alpha$ is a tunable parameter as in traditional Gaussian smearing.
Such a momentum source is designed to increase the overlap with nucleons of the desired boost momentum, and we are able to reach higher boost momentum for the nucleon states than our previous work~\cite{Chen:2017mzz}.

\begin{table*}[htbp!]
\begin{center}
\begin{ruledtabular}
\begin{tabular}{l|cccccccccc}
Ensemble ID & $a$ (fm) & $N_s^3 \times N_t$& $M_\pi^\text{val}$ (MeV) & $M_\pi^\text{val} L$ & $t_\text{sep}/a$ & $P_z$ & $N_\text{cfg}$ & $N_\text{meas}$  \\
\hline
a12m310  & 0.1207(11) & $24^3\times 64$ & 310(3)  & 4.55 & $\{6,7,8,9\}$      &  $\{3,4,5\}\frac{2\pi}{L}$  & 909  &  $\{18180, 29088, 43632, 50904\}$   \\
a12m220S & 0.1202(12) & $24^3\times 64$ & 225(2)  & 3.29 & $\{6,7,8,9\}$      &  $\{4,5,6\}\frac{2\pi}{L}$ &  958  & $\{22922,45984,45984,61312\}$  \\
a12m220  & 0.1184(10) & $32^3\times 64$ & 228(2)  & 4.38 & $\{6,7,8,9\}$      &  $\{3,4,5\}\frac{2\pi}{L}$ &  725  & $\{11600,23200,23200,46400\}$  \\
a12m220L & 0.1189(09) & $40^3\times 64$ & 228(2)  & 5.5  & $\{6,7,8,9\}$      &  $\{4,5,6,8,10\}\frac{2\pi}{L}$ &  840  & $\{13440,26800,26800,53760\}$ \\
a09m130  & 0.0871(6)  & $64^3\times 96$ & 138(1)  & 3.90 &  $\{8,9,10,12\}$&  $\{10,12,14 \}\frac{2\pi}{L}$  &  884  &  $\{17680,28288,56576,109616\}$ \\
a06m310  & 0.0582(4)  & $48^3\times 96$ & 320(2)  & 4.52 & $\{10/12,14,16,18\}$  &  $\{4,5,6,7\}\frac{2\pi}{L}$    & 935  & $\{14960,29920,59840,89760\}$ \\
\end{tabular}
\end{ruledtabular}
\end{center}
\caption{\label{tab:params}
Ensemble information and parameters used in this calculation.
$N_\text{meas}$ is the total number of measurements of the three-point correlators for different values of $t_\text{sep}$.
$L$ indicates the spatial length which is $aN_s$ (in fm).
}
\end{table*}

On the lattice, we first calculate the time-independent, nonlocal (in space, chosen to be the $z$ direction) correlators of a nucleon with finite-$P_z$ boost
\begin{equation}
\label{eq:qlat}
\tilde{h}_\text{lat}(z,P_z) =
  \left\langle \vec{P} \right|
    \bar{\psi}(z) \Gamma \Big( \prod_n U_z(n\hat{z})\Big) \psi(0)
  \left| \vec{P} \right\rangle,
\end{equation}
where $U_z$ is a discrete gauge link in the $z$ direction and $\vec{P}=\{0,0,P_z\}$ is the momentum of the nucleon.
$\Gamma=\gamma^t$ for the unpolarized parton distribution.
Note that our previous work on the unpolarized quark distribution uses $\Gamma=\gamma^z$;
this operator has mixing with matrix elements with $\Gamma=1$~\cite{Constantinou:2017sej,Chen:2017mie}, while the $\gamma^t$ case is free from such mixing at $O(a^0)$.
In this work, we only study the isovector unpolarized quark PDF.

As we increase the nucleon boost momentum, we anticipate that excited-state contamination worsens, since the states are relatively closer to each other;
therefore, a careful study of the excited-state contamination is necessary for the LaMET (or quasi-/pseudo-PDF) approach.
To make sure the excited-state contamination is under control, we measure at least four nucleon three-point source-sink separations, and we perform a number of different extraction and analysis schemes.
We use multigrid algorithm~\cite{Babich:2010qb,Osborn:2010mb} in the Chroma software package~\cite{Edwards:2004sx} to speed up the inversion of the quark propagator for the clover fermions.
Details of our calculation parameters can be found in Table~\ref{tab:params}.

\begin{figure*}[htbp]
\includegraphics[width=.40\textwidth]{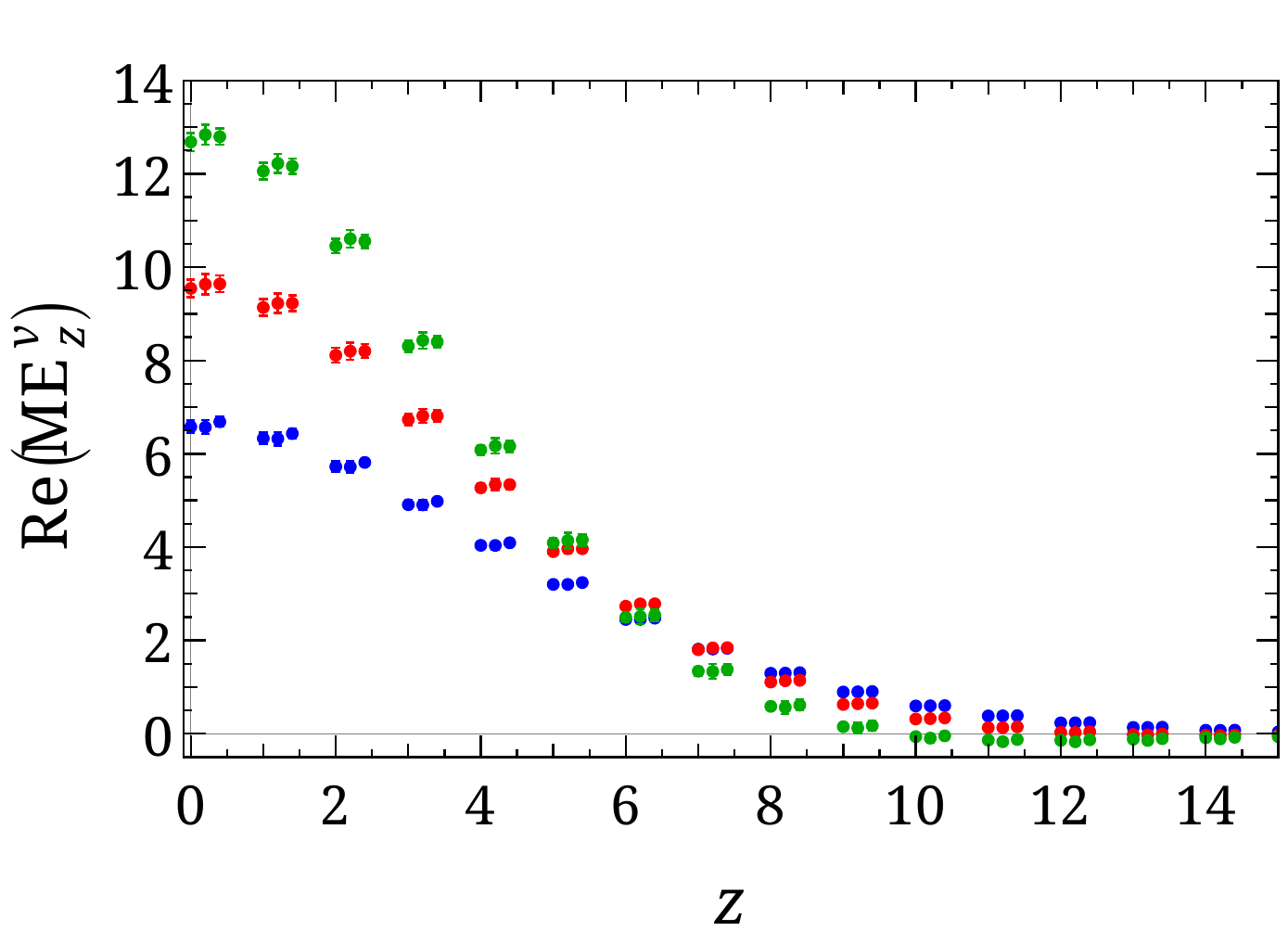}
\includegraphics[width=.40\textwidth]{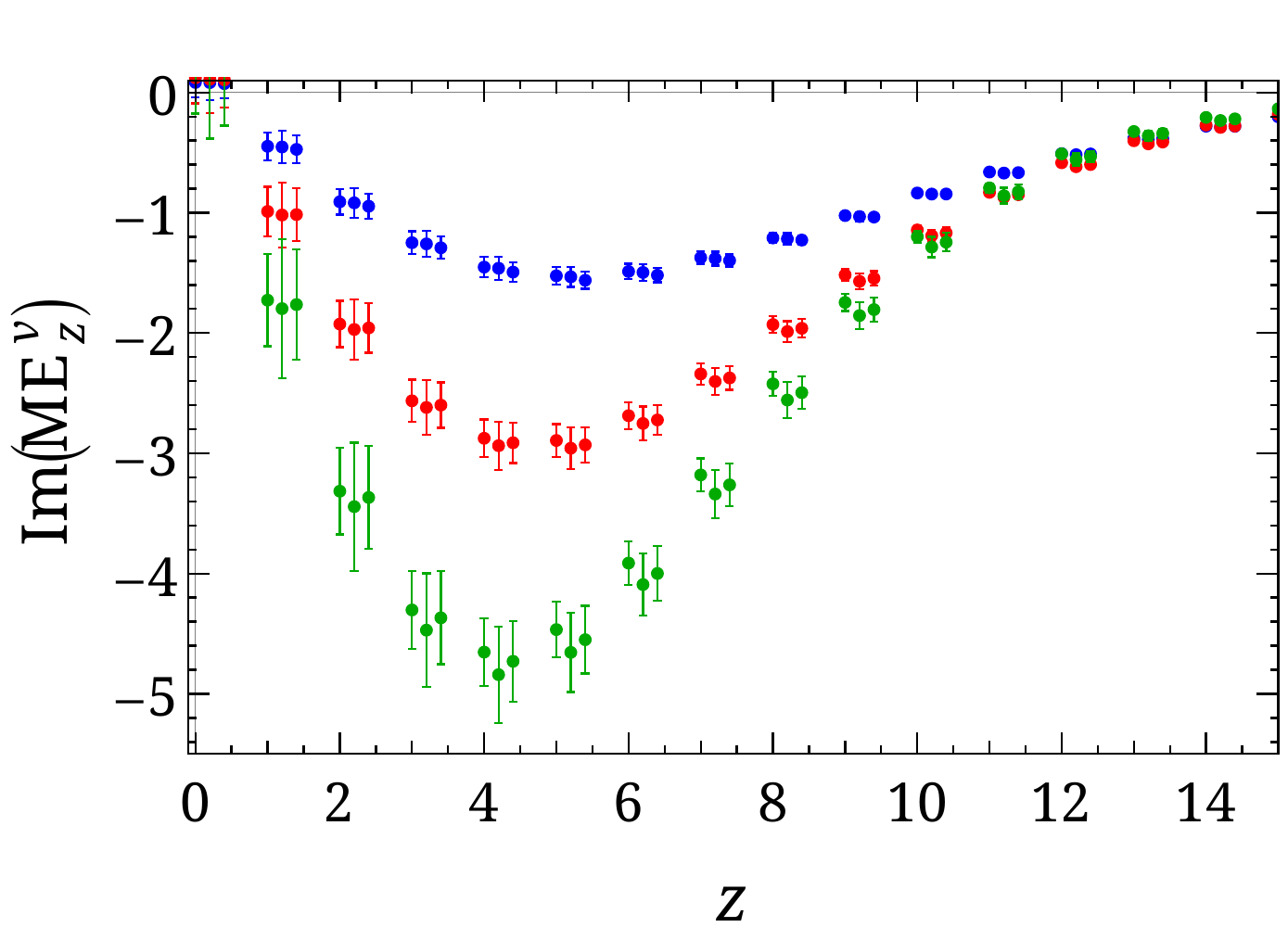}
\caption{The real (left) and imaginary (right) parts of the bare isovector nucleon matrix elements for unpolarized PDFs as functions of $z$ at different momenta.
Their kinematic factors are omitted to enhance visibility by separating the $z=0$ matrix elements.
The colors indicate the different nucleon boost momenta: blue, red and green for matrix elements from 1.7, 2.15 and 2.6~GeV, respectively.
At a given positive $z$ value, the data is slightly offset to show different ground-state extraction strategies;
from left to right they are: two-simRR using all $t_\text{sep}$, two-simRR using the largest 4 $t_\text{sep}$, two-sim using the largest 3 $t_\text{sep}$. Different analyses are consistent within statistical errors, which suggests the excited-state contamination is well controlled.
}
\label{fig:bareME-tsep}
\end{figure*}

Figure~\ref{fig:bareME-tsep} shows an example analysis we did on the ensemble with $a\approx 0.06$~fm and 310-MeV pion mass.
One this ensemble, we use multiple values of nucleon boost momenta, $P_z=\{0,0, n \frac{2\pi}{L}\}$, with $n \in \{4,5,6\}$, which correspond to 1.7, 2.15 and 2.6~GeV nucleon momenta.
We consider multiple analysis methods to remove excited-state systematics among the 5 source-sink separations, 0.60, 0.72, 0.84, 0.96, 1.08~fm, used in this work:
First, we use the ``two-simRR'' analysis described in Ref.~\cite{Bhattacharya:2013ehc} to obtain the ground-state nucleon matrix elements using all five source-sink separations.
(This analysis not only obtains the ground-state matrix element but also the transition and excited-state matrix elements.)
A second extraction uses the same method but only the largest four separations.
Finally, we use the ``two-sim'' analysis, which includes both the ground state and the transition matrix elements but without the excited matrix elements.
Figure~\ref{fig:bareME-tsep} shows the real and imaginary parts of the matrix elements for all three momenta using various combinations of data and analysis strategies.
There is no clear observation of excited-state contamination using any of these analyses.
If the excited states were not under control, we should see these different analyses giving very different ground-state signals.
Similar analysis has been done in all ensembles.
For the rest of this paper, we will take the middle analysis, focusing on the matrix element using ``two-simRR'' with source-sink separation $t_\text{sep} \leq 0.72$~fm only.

\begin{figure*}[htbp]
\includegraphics[width=.4\textwidth]{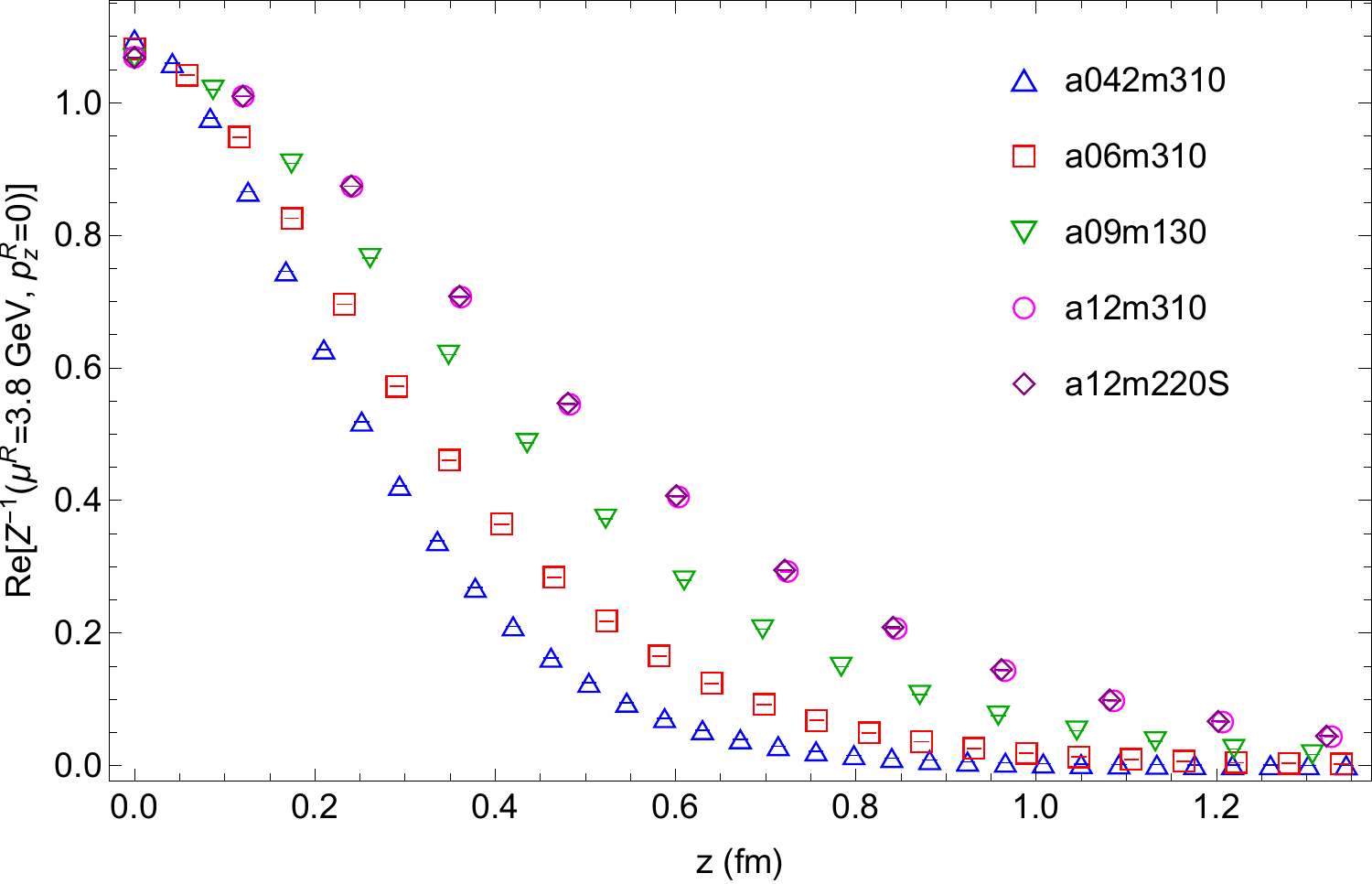}
\includegraphics[width=.4\textwidth]{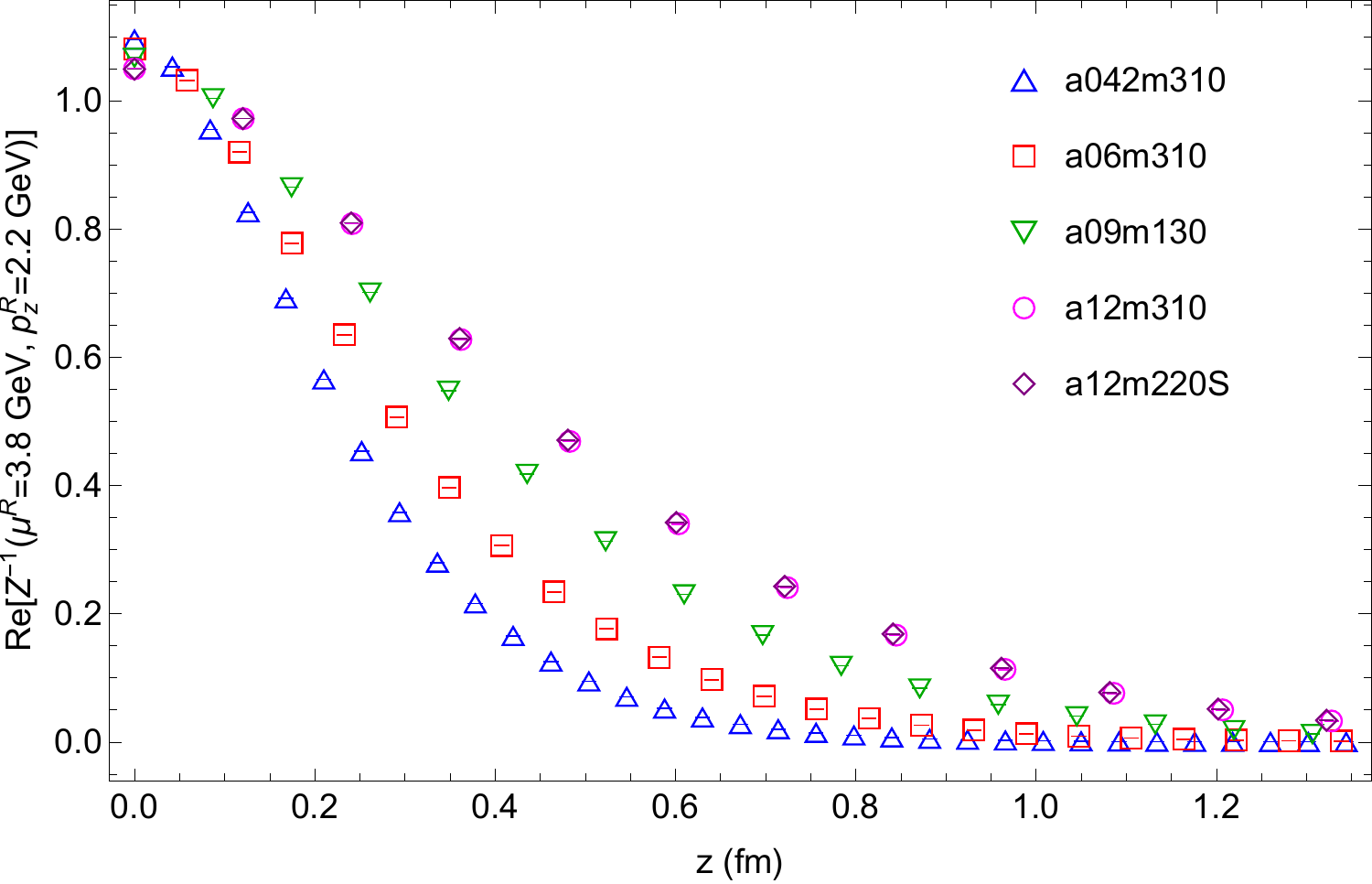}
\caption{
The real component of the inverse renormalization factor from all ensembles as functions of Wilson-line displacement $z$ with RI/MOM renormalization scales $\mu_R =3.8$~GeV and $p_z^R=0$ (left) and $p_z^R=2.2$~GeV (right). 
}
\label{fig:NPR-plots}
\end{figure*}

Before we can study the PDFs, we first need renormalize the bare matrix elements obtained on the ensembles.
To do so, we calculate the RI/MOM renormalization constant $\tilde{Z}$  nonperturbatively on the lattice by imposing the following momentum-subtraction condition on the matrix element of the quasi-PDF in an off-shell quark state:
\begin{multline}\label{hRx}
Z(p^R_z, 1/a, \mu_R) = \\
\left.\frac{\Tr[\slashed p \sum_s \langle ps| \bar\psi_f(\lambda \tilde n) \slashed{\tilde n_t} W(\lambda\tilde n,0) \psi_f(0)|ps\rangle]}
{\Tr[\slashed p  \sum_s \langle ps| \bar\psi_f(\lambda \tilde n) \slashed{\tilde n_t} W(\lambda\tilde n,0) \psi_f(0) |ps\rangle_\text{tree}]} \right|_{\tiny\begin{matrix}p^2=-\mu_R^2 \\ \!\!\!\!p_z=p^R_z\end{matrix}}.
\end{multline}
On the lattice, $\langle ps|O_{\gamma_t}(z)|ps\rangle$ is calculated from the amputated Green function of $O_{\gamma_t}$ with Euclidean external momentum.
In Fig.~\ref{fig:NPR-plots} we show the RI/MOM renormalization factors calculated from all ensembles as a function of Wilson-line displacement $z$.
We observe a strong dependence of the renormalization factors on lattice spacing;
this is expected, since the renormalization factors serve as counterterms to cancel the ultraviolet (UV) divergence of the bare matrix elements.
On the other hand, the dependence on pion mass is negligible;
the renormalization factors from a12m220 and a12m310 overlap one another.

\begin{figure*}[htbp]
\includegraphics[width=.4\textwidth]{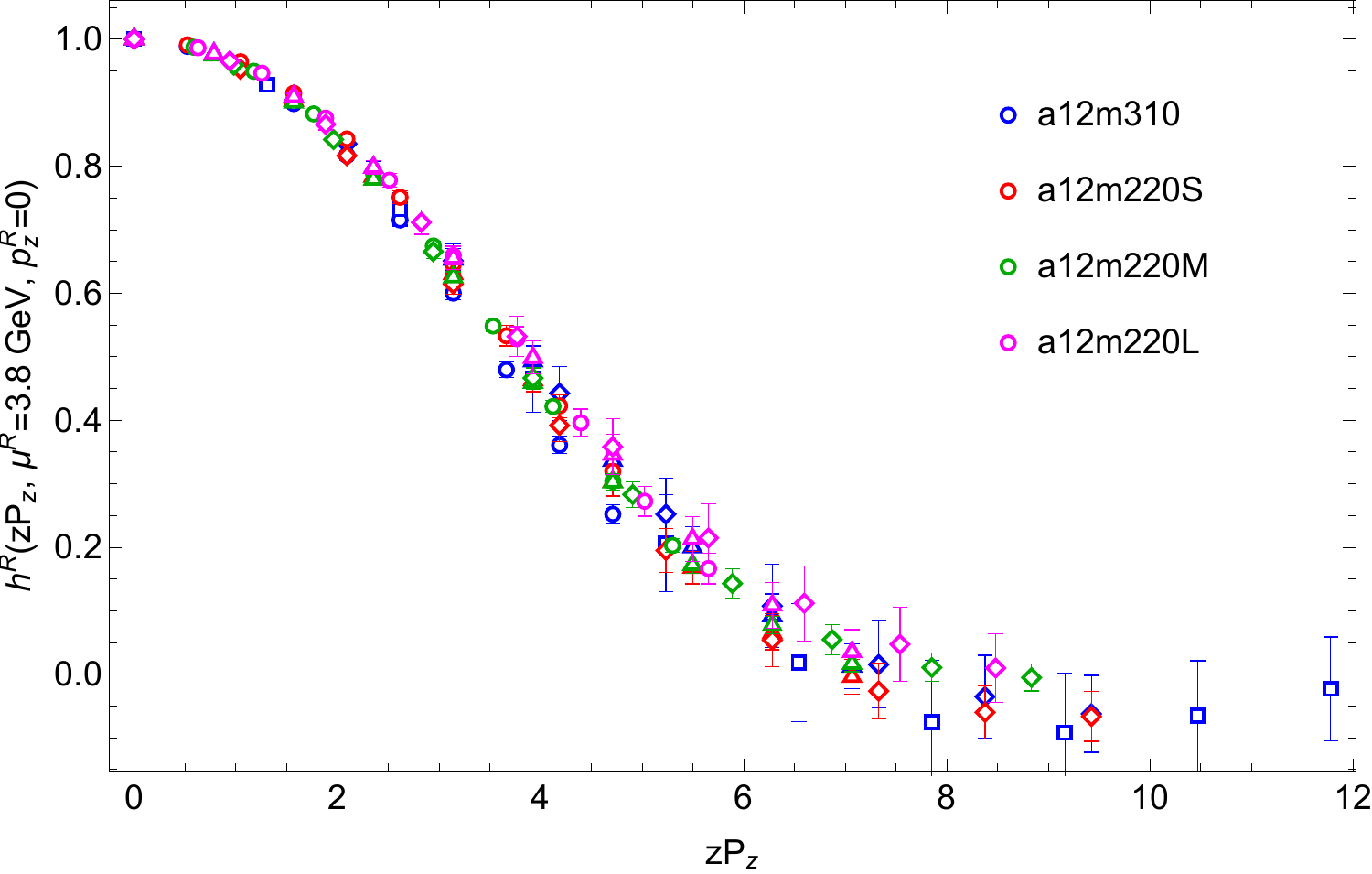}
\includegraphics[width=.4\textwidth]{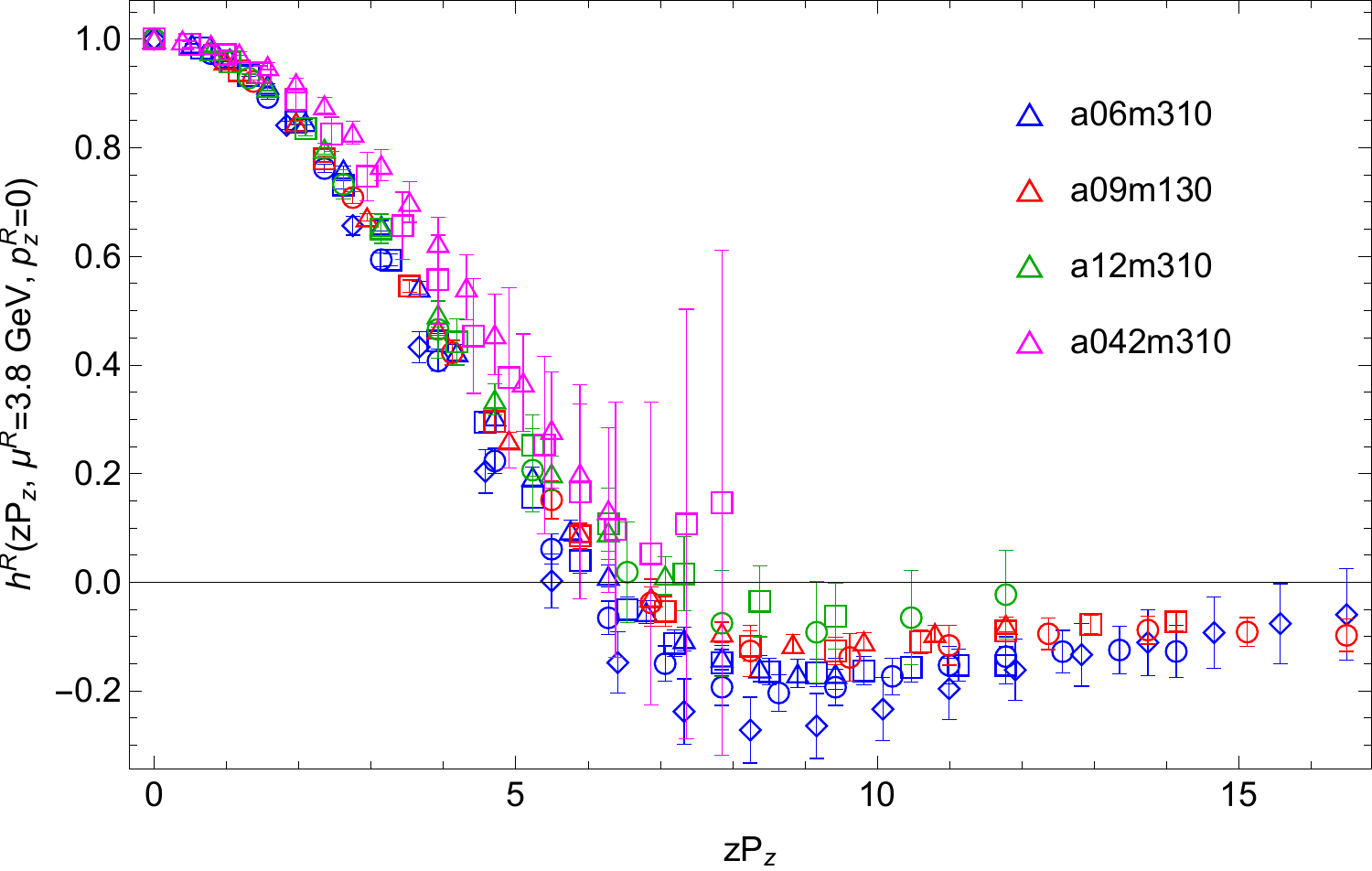}
\caption{
Example of the real renormalized matrix elements with $\mu_R =3.8$~GeV and $p_z^R=0$~GeV comparison of selected $a\approx 0.12$~fm lattices (left) and lattice-spacing dependent 310~MeV results (right) along with physical pion mass ensemble.
The a42m310 data is taken from Ref.~\cite{Fan:2020nzz}, which uses a similar mixed action but much finer lattice spacing.
The lattice ensembles are distinguished by color while different symbols indicate different boost momenta within each ensemble. 
}
\label{fig:RenormME}
\end{figure*}

Figure~\ref{fig:RenormME} shows an example comparison of the real renormalized isovector nucleon matrix elements for all ensembles.
We observe a small pion-mass dependence for the $a\approx 0.12$~fm ensembles between the ensembles with 220- and 310-MeV pions, and no sizable finite-volume effects.
When comparing lattice-spacing dependence, we noted a small trend of the matrix elements moving downward from 0.12~fm to 0.06~fm (green to blue points) but overall within 2 standard deviations.
We also compare with the results from a single superfine lattice-spacing study from Ref.~\cite{Fan:2020nzz} with similar nucleon boost momentum, and the results are consistent as well (due to the larger uncertainties).
For the work below, we will focus on a continuum extrapolation without the superfine lattice spacing, since the data is unlikely change the extrapolation much. 

\section{Results and Discussions}\label{sec:results}

To obtain the physical-continuum matrix elements, we extrapolate the lattice spacing to zero and the pion mass to its physical value through the following ansatz:
\begin{multline}
\label{eq:extrapolation_form}
  h^R(zP_z, a, M_\pi) = \\
  h_0^R(zP_z)\left(1 + c_{a,i}(zP_z) a^i + c_{M,j}(zP_z) M_\pi^j\right)
\end{multline}
To allow for flexibility in the extrapolation form, we vary the order of dependence on lattice spacing ($i$) and pion-mass ($j$) between linear and quadratic.
The finite-volume effects are small in Fig.~\ref{fig:RenormME}, consistent with the ChPT study~\cite{Liu:2020krc}, suggesting that finite-volume effects are negligible for current lattice precision.
In this work, we attempt to use a common set of momenta across as many ensembles as possible and keep any interpolation close to an existing data point.
For this reason, we use $P_z \approx 2.2$ and 2.6~GeV.
Consider $P_z\approx 2.2$~GeV;
this boost momentum corresponds to $n_z=5$ in lattice momentum units on a06m310, a12m310, and a12m220S, $n_z=10$ on a09m130, and $n_z \approx 8.3$ on a12m220L.
Thus, we only need an interpolation of the a12m220L data to get this momentum.
We apply a 3rd-order $\mathbf{z}$-expansion~\cite{Hill:2010yb,Bhattacharya:2011ah} to the matrix elements at five momenta $n_z={4,5,6,8,10}$ on the a12m220L lattice:
$h^R(\mathbf{z},a,M_\pi,L) = \sum_{i=0}^{3} c_{z,i} \mathbf{z}(P_z)^i$, 
then evaluate the polynomial at $P_z = 2.2$~GeV. Because the interpolation is anchored by an existing data point $n_z=8$, we need not worry much about the possibility of overfitting.

\begin{figure*}[htbp]
\includegraphics[width=.45\textwidth]{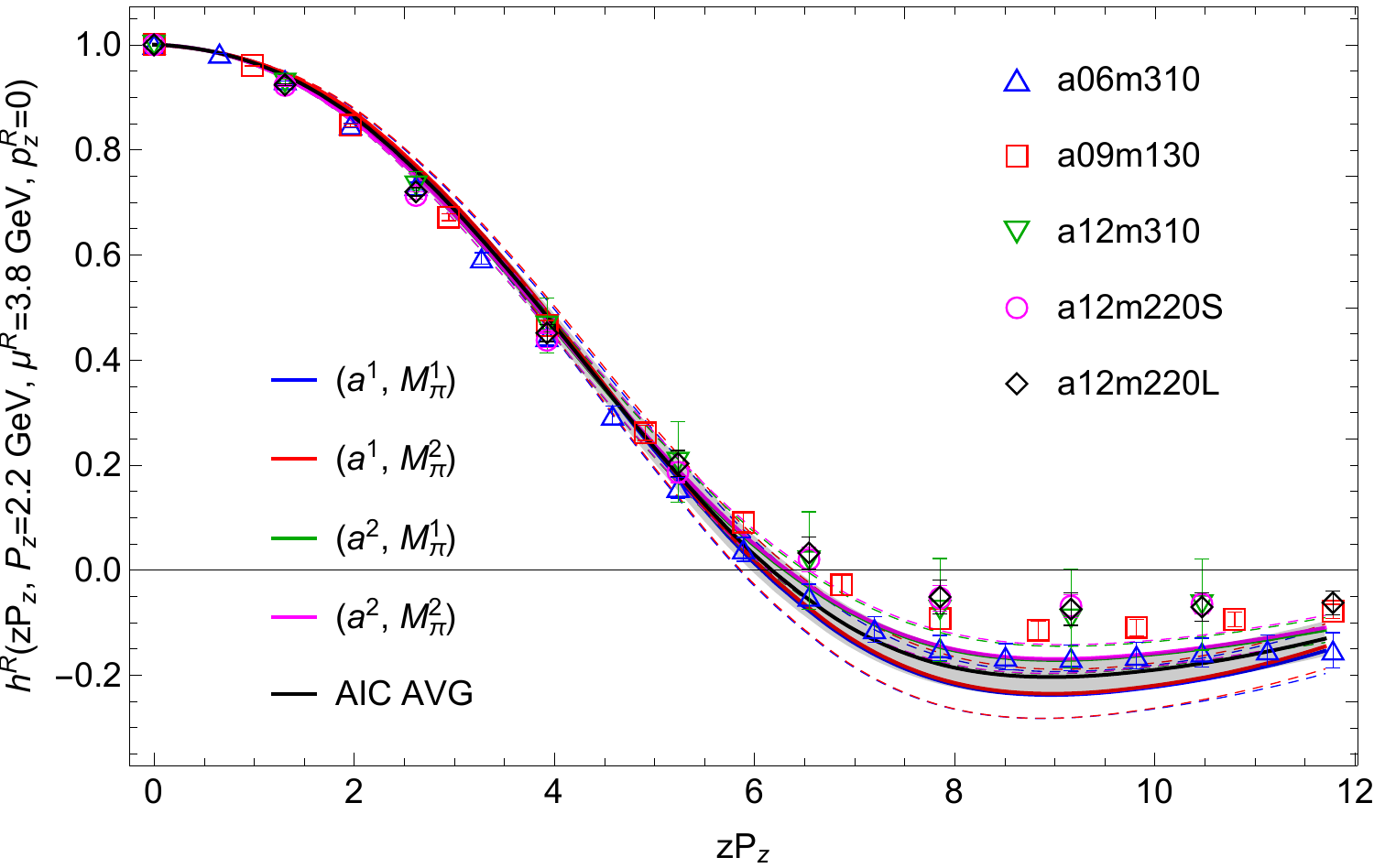}
\includegraphics[width=.45\textwidth]{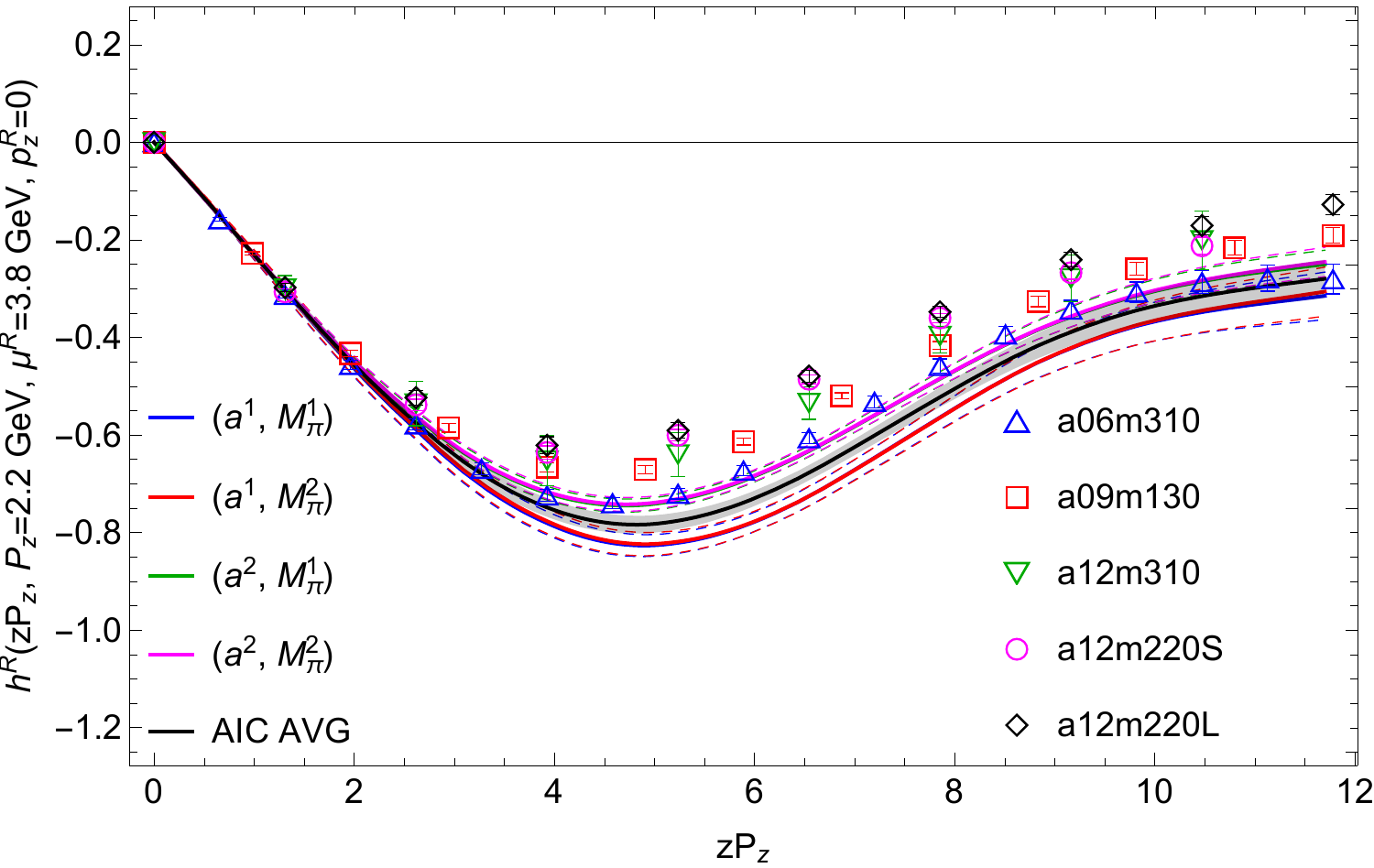}
\caption{
Example of the physical-continuum extrapolation of the real (left) and imaginary (right) matrix elements from the ensembles with nucleon boost momentum around 2.2~GeV.
Various ans\"{a}tze with linear/quadratic extrapolation in lattice spacing and pion mass are shown as solid lines for the central values and dashed lines for uncertainties.
The filled band shows the AIC-averaged physical-continuum matrix elements. 
}
\label{fig:P2GeV}
\end{figure*}

We extrapolate the renormalized matrix elements to the physical limit with four combinations of $i$ and $j$ in Eq.~\eqref{eq:extrapolation_form}, and obtained 4 different physical-continuum matrix elements, as shown in solid (central value) and dashed (error band) lines in Fig.~\ref{fig:P2GeV}.
We find that the four different fits in real matrix elements are in good agreement, and more fluctuations are seen in the imaginary matrix elements.
The fluctuation is mainly dominated by the lattice-spacing extrapolation.
Using a linear lattice-spacing extrapolation form results in slightly higher continuum-limit matrix elements than those obtained from a quadratic form.  

We then average the results the above fits using the Akaike information criterion (AIC):
\begin{equation}
  h^\text{AIC} = \frac{\sum_{i,j}h^{i,j}e^{-(2k_{i,j}+\chi_{i,j}^2)/2}}
                      {\sum_{i,j}e^{-(2k_{i,j}+\chi_{i,j}^2)/2}},
\end{equation}
where $k_{i,j}$ is the number of free parameters to fit, and $\chi_{i,j}^2$ represents the fit quality, which is shown as the gray band in Fig.~\ref{fig:P2GeV}.
The AIC-average results are within two standard deviations of each of the individual fitted matrix elements.
When the nucleon momentum $P_z \gg \{M_N, \Lambda_\text{QCD}\}$, the quasi-PDF can be matched to the PDF through the factorization theorem~\cite{Ji:2013dva,Ji:2014gla,Izubuchi:2018srq},
\begin{equation} \label{eq:fact}
\tilde{q}(x,P_z, p^R_z,\mu_R)=\int_{-1}^1 \frac{dy}{|y|}\: C\left(\frac{x}{ y},r,\frac{yP_z}{\mu},\frac{yP_z}{p_z^R}\right) \, q(y,\mu), 
\end{equation}
where $r=\mu_R^2/{p_z^R}^2$ and $C$ is a perturbative matching kernel, which has been used in previous works~\cite{Chen:2018xof,Lin:2018qky,Chen:2018fwa,Chen:2019lcm}. 
The flavor indices of $q$, $\tilde{q}$, and $C$ are implied.

There are two main sources of residual $P_z$ dependence in removing the frame dependence from the lightcone PDFs: the target-mass correction and twist-4 effects.
For the former, the nucleon mass ($M_N$) corrections can be corrected to all orders in $M_N/P_z$~\cite{Chen:2016utp}.
The twist-4 effect is $\mathcal{O}(\Lambda_\text{QCD}^2/P_z^2)$ from dimensional analysis;
however, Ref.~\cite{Braun:2018brg} suggested the effect could be up to $\mathcal{O}(\Lambda_\text{QCD}^2/x^2P_z^2)$ in order to cancel the renormalon ambiguity in the kernel.
However, a recent study of bubble-chain diagrams in Ref.~\cite{Ji:2020brr} did not find slow convergence of the kernel at three-loop order, indicating that the renormalon effect could be mild to this order in quasi-PDFs.

A related issue is that
Ref.~\cite{Rossi:2018zkn} asserted that the twist-4 operator is set by the lattice spacing $a$;
hence, its suppression factor compared with twist-2 is $\mathcal{O}(1/(P_z a)^2)$ instead of $\mathcal{O}(\Lambda^2_\text{QCD}/P_z^2)$~\cite{Rossi:2018zkn}.
However, the twist-4 contribution that needs to be subtracted from the quasi-distribution operator can be written as an equal-time correlator with two more mass dimensions than the original~\cite{Chen:2016utp}.
Hence, they should not cause power-divergent mixings that need to be subtracted before applying RI/MOM renormalization.
Furthermore, the RI/MOM renormalization factor $Z$ is well fitted by $e^{(m_{-1}/a-m_0)|z|} |z|^{d_1}/c1$ when $z \gg a$~\cite{Ji:2020brr}.
If a power divergence appeared, $Z$ would have extra powers of $1/a$ dependence, which are not observed.

Recently, it was argued that in addition to the typical $\mathcal{O}(1/P_z^2)$ power corrections, nonperturbative renormalization could introduce an even more important $\mathcal{O}(1/P_z)$ infrared contribution that cannot be removed by the matching kernel~\cite{Ji:2020brr}.\footnote{This is because in the $e^{(m_{-1}/a - m_0)|z|}$ structure of the renormalization factor, $m_{-1}$ does not depend on the matrix element used for its extraction, but $m_{0}$ does. It was then shown that this uncertainty induced an $\mathcal{O}(1/P_z)$ uncertainty in the quasi-PDFs~\cite{Ji:2020brr}.}
Hence, we will use this more conservative estimate for our error.

\begin{figure*}[htbp]
\includegraphics[width=.45\textwidth]{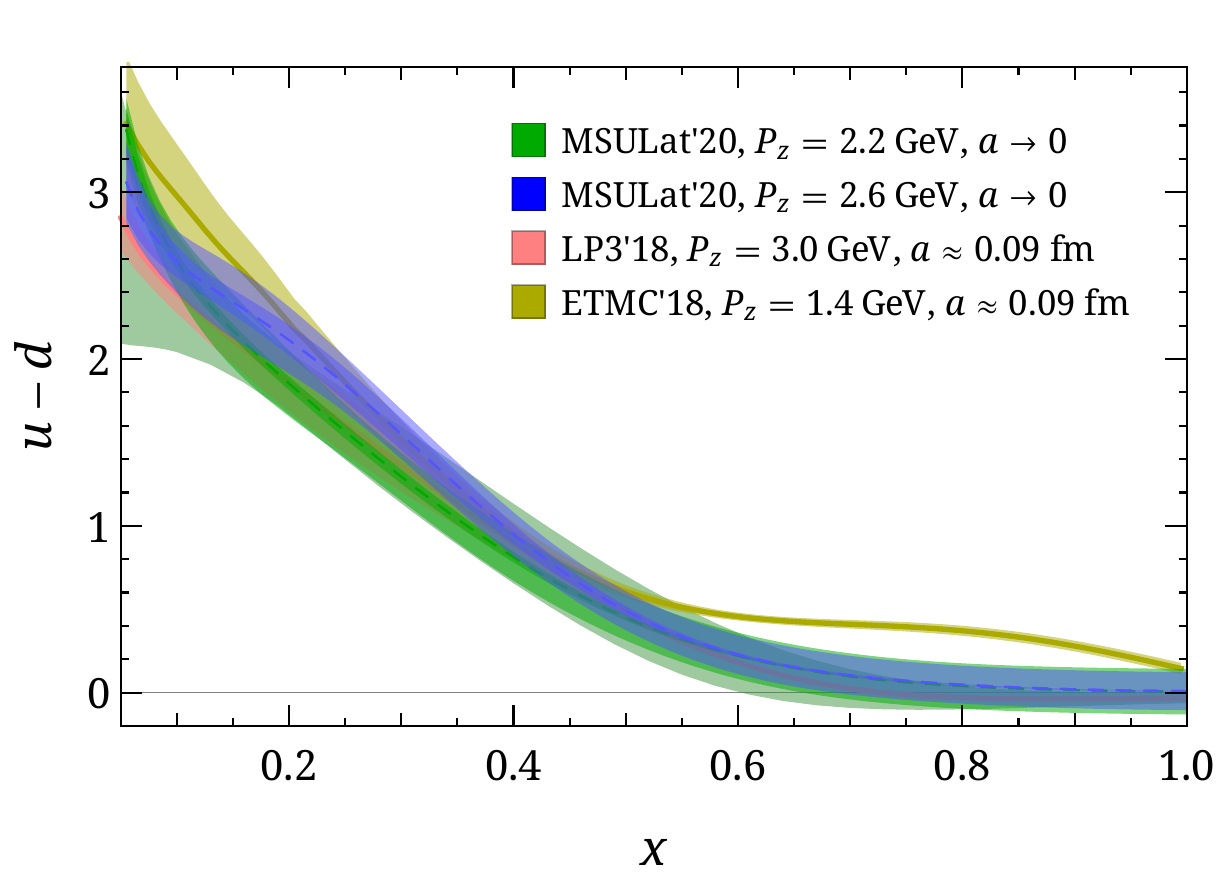}
\includegraphics[width=.45\textwidth]{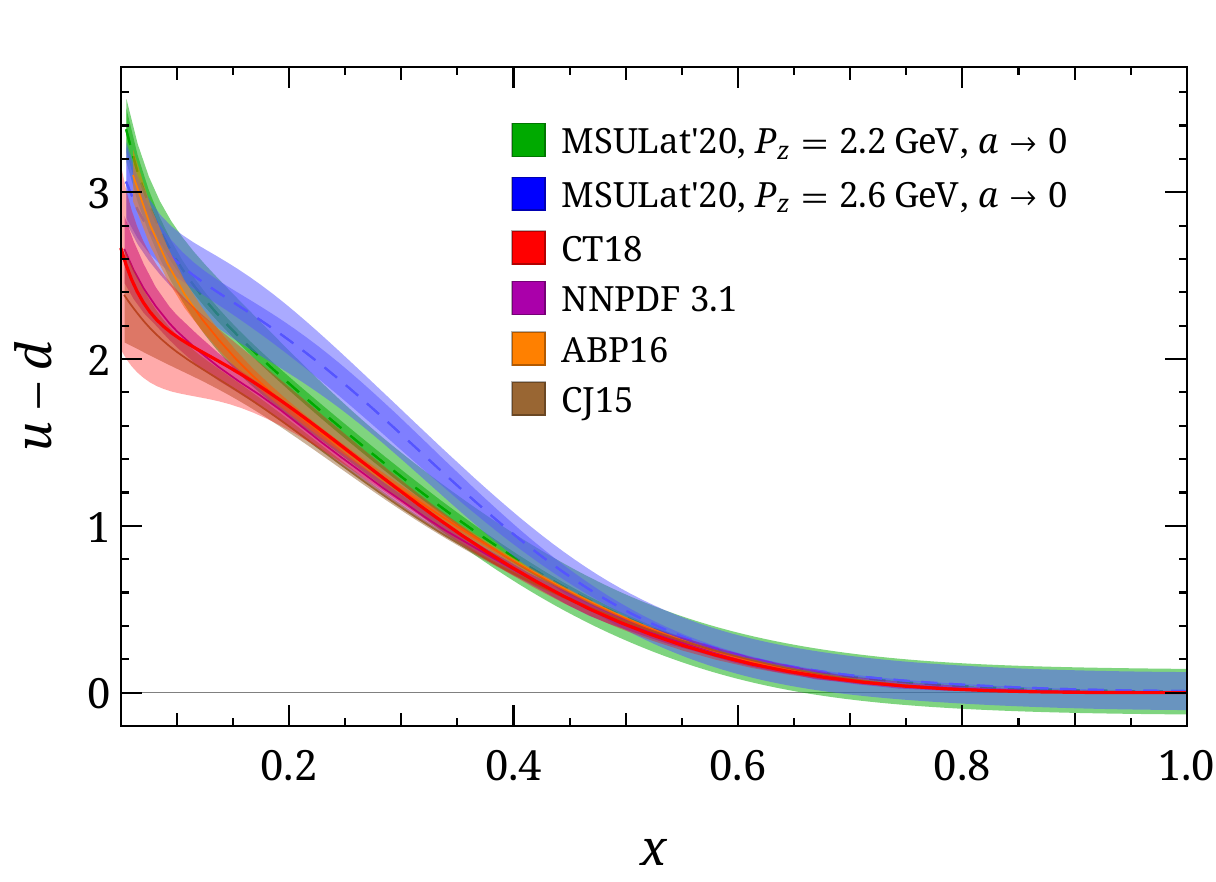}
\caption{
The nucleon isovector unpolarized PDFs from our lattice calculation in the physical-continuum limit, compared with past lattice quasi-PDF results from LP$^3$ and ETMC (left)~\cite{Chen:2018xof,Alexandrou:2018pbm}, 
and global fits from Refs.~\cite{Hou:2019efy,Alekhin:2017kpj,Ball:2017nwa,Accardi:2016qay} (right). 
Note that the previous lattice work by LP3'18 and ETMC'18 were done using a single lattice spacing at physical pion mass and did not take into account the systematics due to twist-4 effects, while our work (MSULat'20) includes this systematic as well as the reconstruction errors.
}
\label{fig:PDFcomp}
\end{figure*}

We estimate the systematic associated with the transformation of the lightcone distribution through the following procedure.
First, we take the nucleon isovector PDF from CT18 global fit, and create set of mock matrix elements as functions of $zP_z$ using the same parameters used in the lattice calculation.
We then run these mock matrix elements through the same analysis used to calculate the PDFs;
this should yield the same PDFs that were originally used to create the data, but they will differ due to the inverse problem in the transformation.
A similar analysis has been done in Ref.~\cite{Lin:2017ani}.
The difference between the input and reconstructed PDFs provides a measure of the size of the transformation systematic uncertainty.
As expected, the reconstructed PDFs have much larger uncertainties in the small-$x$ and negative-$x$ regions.
We neglect the small-$x$ and antiquark results due to the large uncertainty associated with nucleon boosted momenta less than 2.6~GeV.
We added the difference as a systematic error in quadrature with the twist-4 errors, estimated to be $\mathcal{O}(\Lambda_\text{QCD}/P_z)$ by using $\Lambda_\text{QCD}\approx 0.3$~GeV\footnote{The error coming from $\mathcal{O}(\Lambda_{QCD}^2/P_z^2)$ is only a few percent, too small to be seen at the scale of the results, so we ignore it here and focus on the larger sources of uncertainty.}.
These errors are shown as the outer uncertainty band in Fig.~\ref{fig:PDFcomp}.

We focus on comparing our results with previous lattice quasi-PDF calculations done at the physical pion mass (but with a single lattice spacing) and with a selection of global-fit PDFs.
The first generation of unpolarized PDFs at the physical pion mass~\cite{Lin:2017ani,Alexandrou:2018pbm} using the quasi-PDFs approach were determined using small momentum with $P_\text{max} \approx 1.3$~GeV at a single lattice spacing ($a \approx 0.09$~fm).
This, in addition to the challenges in reconstructing the $x$ dependence, were shown to have led to the wrong sign of sea-flavor asymmetry~\cite{Lin:2017ani}.
Later calculations at physical pion mass pushed the nucleon boost momentum 3~GeV~\cite{Chen:2018xof}.
However, the lattice discretization systematics were not taken into account, and the twist-4 effects were assumed to be $\mathcal{O}(\Lambda_\text{QCD}^2/P_z^2)$.
The latter estimated systematic is negligible, since these few-percent effects at this large momentum are much smaller than the statistical and other systematics.
Since we account for all these neglected systematics in this work, the total uncertainty appears larger than those of previous quasi-PDF works, even though the statistical error remains comparable.
When comparing our continuum-physical nucleon isovector PDFs with those obtained from global fits,
CT18NNLO~\cite{Hou:2019efy},
NNPDF3.1NNLO~\cite{Ball:2017nwa},
ABP16~\cite{Alekhin:2017kpj}, and
CJ15~\cite{Accardi:2016qay},
we found our results, even with only the errors considered by inner statistical bands, have nice agreement.
The errors increase toward the smaller-$x$ region for both lattice and global fitted PDFs, but overall, they agree within two standard deviations.

\section{Summary and Outlook}\label{sec:summary}

In this work, we presented the first determination in the physical-continuum limit of the nucleon isovector parton distribution, using six lattice ensembles, including 3 lattice spacings, multiple volumes and a physical pion mass.
We found small a small pion-mass dependence and no sizable finite-volume effects, but a noticeable trend of the matrix elements changing from 0.12~fm to 0.06~fm.
The resulting continuum-physical matrix elements are dominated by the lattice-spacing extrapolation.
Our analysis results in PDFs consistent with various global PDF fits with excellent agreements for mid- to large-$x$ regions, and compatible within 2 standard deviations for $x < 0.4$.
The nucleon isovector moments $\langle x^n \rangle$ are around 0.2, 0.06, and 0.04 for $n=1,2,3$, respectively.
Currently, we use a conservative systematic error estimate, mainly dominated by twist-4 systematics on the order of $\mathcal{O}(\Lambda_\text{QCD}/P_z)$.
The small-$x$ and antiquark PDFs are not reliably extracted in this work;
future work will focus on reducing the twist-4 systematics and pushing toward improving the lattice determination of small-$x$ and antiquark PDFs.

\section*{Acknowledgments}
We thank the MILC Collaboration for sharing the lattices used to perform this study. The LQCD calculations were performed using the Chroma software suite~\cite{Edwards:2004sx}  with the multigrid solver algorithm~\cite{Babich:2010qb,Osborn:2010mb}.
This research used resources of
the National Energy Research Scientific Computing Center, a DOE Office of Science User Facility supported by the Office of Science of the U.S. Department of Energy under Contract No. DE-AC02-05CH11231 through ERCAP and ALCC;
the Extreme Science and Engineering Discovery Environment (XSEDE), which is supported by National Science Foundation grant number ACI-1548562;
facilities of the USQCD Collaboration, which are funded by the Office of Science of the U.S. Department of Energy,
and supported in part by Michigan State University through computational resources provided by the Institute for Cyber-Enabled Research (iCER).
The work of HL is also partly supported by the  Research  Corporation  for  Science  Advancement through the Cottrell Scholar Award.
HL and RZ are supported by the US National Science Foundation under grant PHY 1653405 ``CAREER: Constraining Parton Distribution Functions for New-Physics Searches''.
JWC is partly supported by the Ministry of Science and Technology, Taiwan, under Grant No. 108- 2112-M-002-003-MY3 and the Kenda Foundation.

\ifx\@bibitemShut\undefined\let\@bibitemShut\relax\fi
\makeatother


\end{document}